\documentclass[pre,twocolumn,showpacs,showkeys]{revtex4}

\usepackage{graphicx}
\usepackage{dcolumn}
\usepackage{bm}
\usepackage{subfigure}
\usepackage{amssymb}
\usepackage{graphics}
\usepackage{hyperref}
\usepackage{epsfig}
\usepackage{threeparttable}

\usepackage{mathrsfs}

\begin{document}
\title{Hysteresis features of the transition-metal dichalcogenides VX$_2$ (X=S, Se, and Te)}
\author{E. Vatansever$^1$}\email{erol.vatansever@deu.edu.tr}
\author{S. Sarikurt$^1$}
\author{R.F.L. Evans$^2$}
\affiliation{$^1$Department of Physics, Dokuz Eyl\"{u}l University, TR-35160, Izmir-Turkey}
\affiliation{$^2$Department of Physics, University of York, YO10 5DD, York-United Kingdom}

\date{\today}
\begin{abstract}
Very recently, it has been shown that vanadium dichalcogenides (VX$_2$, X=S, Se and Te) monolayers 
show intrinsic ferromagnetism, and their critical temperatures are nearly to 
or beyond room temperature.  Hence,  they  would have wide  potential applications in next-generation 
nanoelectronic and spintronic devices. In this work, being inspired by a recent 
study we systematically perform Monte Carlo simulations based on single-site update Metropolis algorithm to investigate the hysteresis features of  VX$_2$ monolayers for a wide range of temperatures up to 600 K. Our simulation results indicate that, both remanence and coercivity values tend to decrease with increasing temperature. Furthermore, it is found that hysteresis curves start to evolve from rectangular at the lower temperature regions to nearly S-shaped with increasing temperature.
\end{abstract}
\maketitle
\section{Introduction}\label{Introduction}
In recent years, two-dimensional (2D) monolayers such as boron nitride (BN)~\cite{han2008structure}, silicene~\cite{vogt2012silicene}, germanene~\cite{davila2014germanene}, arsenene and antimonene~\cite{zhang2015atomically}, phosphorene~\cite{liu2014phosphorene} and transition metal dichalcogenides (TMDs)~\cite{novoselov2005two, coleman2011two, chhowalla2013chemistry} have attracted considerable attention beyond graphene due to their remarkable structural, electronic, optical and magnetic properties. Among these 2D materials, TMDs have extensive and growing application areas ranging from electronics to energy storage~\cite{appel2005molybdenum, karunadasa2012molecular, chang2011cysteine, radisavljevic2011single,wang2012electronics, bhimanapati2015recent}.

Most of the primitive TMDs are non-magnetic and these materials could gain magnetic properties by applying strain or introducing transition metal atoms, point defects or non-metal element adsorption~\cite{shidpour2010density,he2010magnetic,wang2011ultra,li2008mos2,ma2011graphene,ma2011electronic}. In recent years, the effect of tensile strain on magnetic moments of monolayered MX$_2$ (M=V, Nb; X = S, Se) structures have been demonstrated by several researchers~\cite{ma2012evidence, zhou2012tensile, kan2015density}. These studies pointed out that electronic and magnetic properties can be manipulated by applying a biaxial compression or tensile strain. Due to existing intrinsic ferromagnetism and semiconducting properties in VX$_2$ (X = S, Se and Te) monolayer structures,  it's not necessary to expose these materials under the effect of tensile strain or to doping of the monolayer structure with either transition metals or non-metallic atoms~\cite{gao2013ferromagnetism,FuhVX2_2016}. So, this provides us that ferromagnetic TMDs such as pristine VX$_2$ monolayers could be fabricated without either applying strain or introducing transition metal atoms or native defects. 

In the 1970s, bulk VS$_2$ was synthesized for the first time~\cite{murphy1977preparation}. A few-layer VS$_2$~\cite{feng2011metallic, coleman2011two, feng2012giant, gao2013ferromagnetism, zhang2013dimension,rout2013synthesis, song2014highly} has been synthesized in both hexagonal (H-VS$_2$) and trigonal (T-VS$_2$) structures while the bulk and few-layer VSe$_2$~\cite{boscher2007atmospheric,guzman1997vse2,thompson1980electrochemical} and bulk VTe$_2$~\cite{vinokurov2009thermodynamic} can only be successfully synthesized in the trigonal structure. As far as we know, the synthesis of monolayer VX$_2$ (X=S, Se and Te) has not been experimentally demonstrated to date. 

Electronic and magnetic properties of VX$_2$ monolayers (X = S, Se and Te) have been investigated in numerous studies using first-principles calculations based on density functional theory (DFT)~\cite{feng2011metallic, ma2012evidence, gao2013ferromagnetism,zhang2013vanadium,zhang2013dimension,gan2013two,pan2014magnetic,zhong2014ferromagnetism,pan2014electronic, wasey2015quantum,qu2015effect,kan2015density,huang2015prediction, FuhVX2_2016,zhuang2016stability,wang2016magnetic,luo2017structural}. For example, Ma et al.~\cite{ma2012evidence} reported that pristine 2D VX$_2$ (X = S, Se) monolayers exhibit magnetic ordering and that they have two stable magnetic structures, one being ferromagnetic (FM) while the other is antiferromagnetic (AFM). They performed calculations to obtain the magnetic behaviour of the ground state for VX$_2$ monolayers and found out that the ground states of both structures are FM. This ground state characteristic has been explored in various studies~\cite{ma2012evidence,pan2014electronic,wasey2015quantum, FuhVX2_2016, zhuang2016stability}.

According to the results of first-principles calculations, it has been previously stated that the total magnetic moment has the main contribution from V atoms while X atoms only give a small contribution to the magnetism of VX$_2$ (X = S, Se). In the case of VS$_2$; the magnetic moment values have been obtained as $0.486\ \mu_B$~\cite{ma2012evidence}, $0.51\ \mu_B$~\cite{gao2013ferromagnetism}, $0.858\ \mu_B$~\cite{pan2014electronic} on each V atom and $-0.026\ \mu_B$~\cite{ma2012evidence}, $0.03\ \mu_B$~\cite{gao2013ferromagnetism}, $0.047\ \mu_B$~\cite{pan2014electronic} on each $S$ atom. In the VSe$_2$ and VTe$_2$ case, the magnetic values have been reported as $0.680\ \mu_B$~\cite{ma2012evidence}, $0.951\ \mu_B$~\cite{pan2014electronic} per V atom, $0.048\ \mu_B$~\cite{ma2012evidence}, $0.062\ \mu_B$~\cite{pan2014electronic} per Se atom, and $0.986\ \mu_B$~\cite{pan2014electronic} per V atom,  $0.096\ \mu_B$~\cite{pan2014electronic} per Te atom, respectively.

Gao et al.~\cite{gao2013ferromagnetism} have experimentally demonstrated that ultrathin VS$_2$ nanosheets with less  than five layer represent room temperature FM behavior combined with weak antiferromagnetism.  According to the results obtained from the experiment performed at room temperature, the nanosheet demonstrates clear hysteresis in the low applied field whereas it shows paramagnetic behaviour in the high field regime. They also carried out magnetization versus magnetic field (M-H) curve measurement for temperature range from 50 K to 300 K and  for all different temperature values they observed the pronounced S-shaped curve, which is a characteristic of ferromagnetism~\cite{gao2010defect,wang2012magnetic}. They revealed that the saturation magnetization and coercive field decrease with increase of temperature. Furthermore, they verified their experimental results that point out the ferromagnetic ground state by performing total energy calculations. 

Kan et al.~\cite{kan2015density} studied the comparative stability of both H- and T-structures of VS$_2$ monolayer using DFT with generalized gradient approximation (GGA) functionals. They explored that H-structure has a indirect semiconducting character with a small band gap of $0.187$ eV whereas T-structure has a metallic character~\cite{ma2012evidence, feng2011metallic, zhang2013dimension, ataca2012stable} and H-structure is more stable than the T-structure. They also obtained the semiconductor character of H-structure of VS$_2$ monolayer using GGA plus on-site Coulomb interaction U (GGA+U, $U=3$ eV) and Heyd--Scuseria--Ernzerhof (HSE06) hybrid functional. The band gap energies were calculated as  E$_{\text{gap}}=0.721$ eV and E$_{\text{gap}}=1.128$ eV in respective order. Besides these results, Kan et al. also pointed out that both structures are magnetic and T-structure of VS$_2$ monolayer has lower magnetism ($0.43\ \mu_B$  per unit cell) in comparison with H-structure ($1\ \mu_B$ per unit cell). In a previous study by Zhang et al.~\cite{zhang2013dimension}, for the bulk structures the total magnetic moments of hexagonal and trigonal VS$_2$ were estimated as $0.85\ \mu_B$ and $0.31\ \mu_B$, respectively. Due to the difference between magnetic moment values of T and H-structures, it can be deduced that both monolayer and bulk H-VS$_2$ structures exhibit greater magnetism than T-structures  and so the magnetic moment significantly depends on the crystal structure. More recently, Fuh et al.~\cite{FuhVX2_2016} reported that H-VX$_2$ monolayer structures have indirect energy band gap with respective small band gap energies of $0.05, 0.22,$ and $0.20\ eV$ using GGA DFT functionals. They also considered different exchange-correlation functionals such as GGA+U ($U=2$ eV) and HSE06 hybrid functional to examine the electronic band structure of H-VX$_2$ monolayers. In addition, they performed GW calculation method to get an accurate value of the semiconducting band gap. The respective energy band gaps were reported as E$_{\text{gap}}=1.334, 1.2, 0.705$ eV for VS$_2$, VSe$_2$ and VTe$_2$.  Furthermore, Huang et al.~\cite{huang2015prediction} and Zhuang et al.~\cite{zhuang2016stability} performed electronic band structure calculations for H-VS$_2$ monolayer by using GGA+U ($0\leq U \leq 4$ eV) and Local Density Approximation (LDA)+U ($0\leq U \leq 5$ eV), respectively and HSE06 functionals. They all found out that H-VX$_2$ monolayer structures are semiconductors. 

On the other hand, Fuh et al.~\cite{FuhVX2_2016} also performed Monte Carlo simulations and obtained the 
Curie-temperatures as $292$, $472$, and $553$ K for VS$_2$, VSe$_2$, and VTe$_2$ monolayers, respectively. The Curie temperature values obtained using mean field theory are reported by Pan~\cite{pan2014electronic} as follows: $309$ K for VS$_2$, $541$ K for VSe$_2$ and $618$ K for VTe$_2$ monolayers. These Curie temperatures revel that the unique ferromagnetic character can be achieved close to or well beyond the room temperature.

In this paper, we examine the temperature dependence of remanence magnetization and coercivity of H-VX$_2$ (X=S, Se, and Te) monolayers using Monte Carlo simulation with the Metropolis algorithm. We also  evaluate the impact of the temperature on hysteresis loops. 

\section{Model and Simulation Details}\label{Model}
We study hysteresis features of the VX$_2$ (X=S, Se and Te)
monolayers on a hexagonal crystal structure under the influence of a magnetic field. The spin Hamiltonian of the
considered systems can be written as follows:

\begin{equation}\label{Eq1}
\begin{array}{cc}
 &\mathscr{H}=\displaystyle-J_{1}\sum_{\langle ij \rangle} \mathbf{S_{i}}\cdot\mathbf{S_{j}}-J_{2}\sum_{\langle\langle ij \rangle\rangle} \bold{S_{i}}\cdot\mathbf{S_{j}}\\
 \\
 &\displaystyle-k\sum_{i}\left(S_{ix}^2+S_{iy}^2\right)-g\mu_{B}H\sum_{i}S_{ix}.
\end{array}
\end{equation}

Here, $\bold{S_{i}}$ and $\bold{S_{j}}$ are the classical  Heisenberg spins with unit
magnitude at the site $i$ and $j$ in the system.  $J_{1}$ and $J_{2}$ represent  nearest
neighbour (NN) and next-nearest neighbour (NNN) spin-spin couplings, respectively.  $\langle \cdots \rangle$ and
$\langle\langle \cdots \rangle\rangle$ denote the summation over the NN and NNN spin pairs
through the  system, respectively. $k$ term corresponds to the magnetic
anisotropy energy of  the system. $g, \mu_{B}$ and $H$ are Land\'{e}-g constant,
Bohr magneton and magnetic field terms, respectively.
The first two summations are over the NN and NNN spin pairs while the last two summations are
over all of the lattice points in the system. We follow the  Ref.~\cite{FuhVX2_2016} for the real
values of the exchange couplings and magnetic anisotropy energy. We note that in
Table 2 and 4 of Ref. \cite{FuhVX2_2016}, the authors give the calculated magnetic
anisotropy energies and exchange interaction parameters  of VS$_2$, VSe$_2$ and
VTe$_2$ monolayers,  respectively.

We use Monte Carlo simulation with single site update Metropolis
algorithm \cite{Binder, Newman} to understand the magnetic properties of the system
on a $L \times L$ hexagonal crystal structure. Here, $L$ is the linear  size of the
hexagonal lattice, and it is fixed as $120$ through this work. We select the
boundary  conditions such that they are periodic in all directions. The
simulation process can be briefly summarized as follows. Magnetic hysteresis
curves are generated over the 50 independent sample realizations. In each
sample realization, the simulation starts at high temperature value
using random initial spin configuration. It slowly cooled down until the
temperature reaches a specific temperature value in the presence of a magnetic
field $H=1T$ applied in $x$-direction of the system. Then, using the obtained last spin
configuration as initial configuration, the decreasing branch of
hysteresis curve is obtained by tracing the magnetic field
from  $H$ to $-H$. As the applied field
reaches $-H$ value, it means that the decreasing branch of hysteresis
curve is completed. Similarly,  using the spin configuration obtained after this
process as initial configuration, the increasing branch of
hysteresis curve is obtained by scanning the magnetic field
from  $-H$ to $H$. At each magnetic field step,
the magnetization is collected over $5\times10^4$ Monte Carlo
per step (MCSS) after  the first $10^4$ MCSS are discarded for
equilibration process. The instantaneous magnetization components of
the system can be calculated using the formula $M_{\alpha}=1/N\sum_{i}S_{i\alpha}$. Here, $N$ is
the total number of the spins located on the hexagonal lattice,
and $\alpha=x,y$ and $z$.

\section{Results and Discussion}\label{Discussion}

\begin{figure}[h!]
\center
\includegraphics[width=7cm]{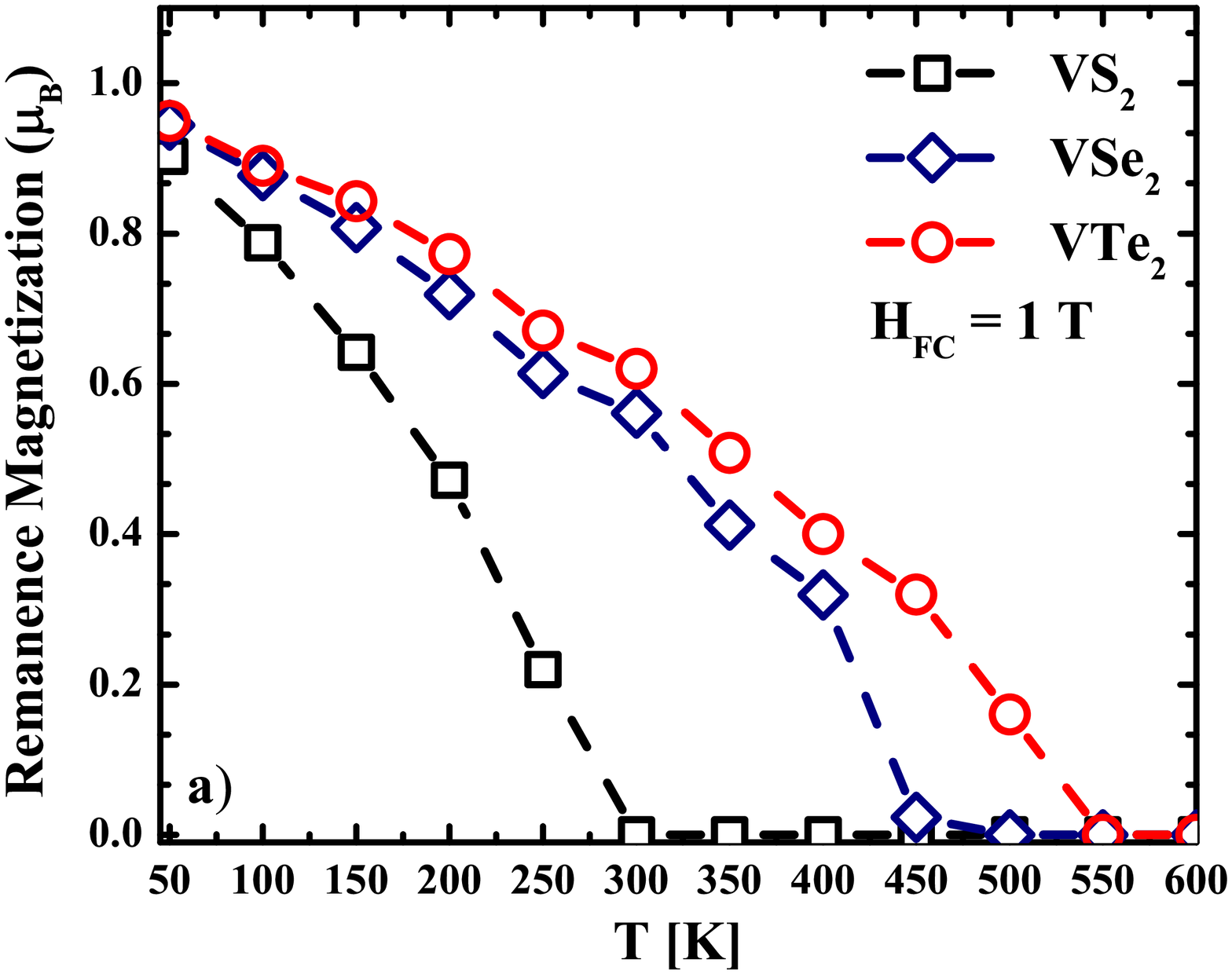}
\includegraphics[width=7cm]{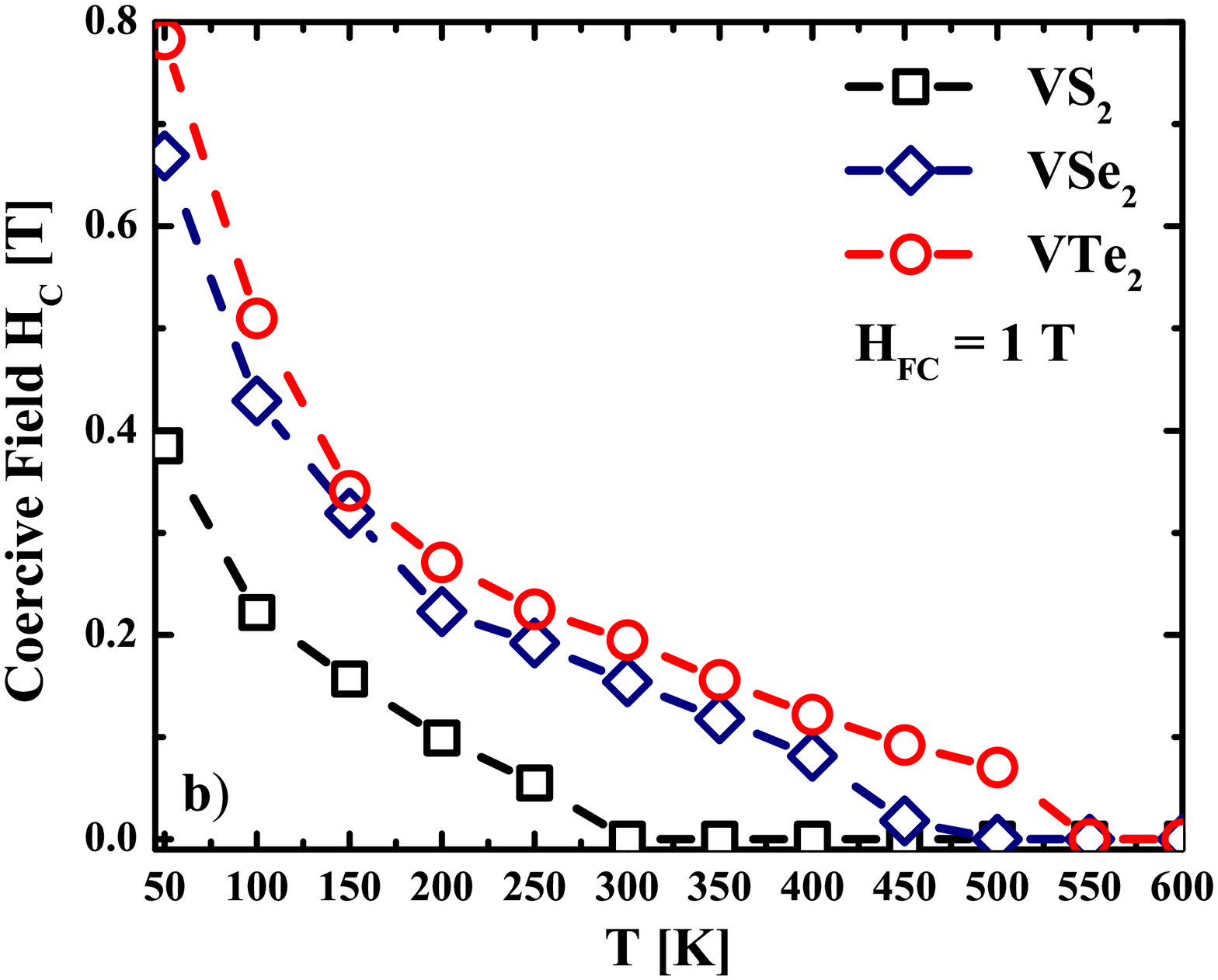}
\caption{(Color online) Variations of the (a) remanence magnetizations 
and (b) coercivities as functions of the temperature of the VX$_{2}$ monolayers.
All curves are measured under the field cooling H$_{FC}= 1 T$ applied in 
parallel directions with respect to the  VX$_2$ monolayers.}\label{Fig1}
\end{figure}

In Fig. \ref{Fig1}, we give thermal dependencies of the remanence magnetizations 
and coercivities of the  VX$_{2}$ monolayers. As shown in Fig. \ref{Fig1}(a), remanence 
magnetization treatments of the considered systems sensitively depend on the studied temperature value. 
In the lower temperature regions, VX$_{2}$ monolayers are in the strongly ferromagnetic phase. So, 
it is possible to observe bigger remanence values in this region. When the temperature is increased  
starting from relatively lower  temperature, remanence magnetizations become to shift to the lower 
values. It means  that increasing thermal  energy tends to overcome the spin-spin interaction term 
between spin pairs and magnetic anisotropy energy in the system. As expected,  if the temperature is 
increased further and is reached to a characteristic value, remanence magnetization tends to vanish.  
When one compares the obtained remanence magnetizations as functions of the temperature for VX$_{2}$ 
monolayers, one can easily see that VTe$_{2}$ has much higher remanence value than the those found in 
other two materials.  As given in Fig. \ref{Fig1}(b) we also  investigated coercivity 
treatments of the VX$_{2}$ monolayers. It is possible to state that
much more energy originating from external field is needed to reverse the sign of magnetization at the 
relatively lower temperature regions where VX$_{2}$ monolayers display strong ferromagnetic character. 
When the temperature is increased starting from lower temperature regions, coercive field values starts to 
decrease. Actually, this is an expected result since an increment in the temperature gives rise to  the 
occurrence of more fluctuating behavior in the system, so coercive field gets narrower with increasing 
temperature value. These comments are also valid for all three  monolayers studied here. If one compares 
the coercivity curves of VX$_{2}$ monolayers, one can see that VTe$_{2}$ has much higher coercivity 
value than the those found in others two materials. It should be mentioned here that according to 
the Ref.~\cite{FuhVX2_2016}, Curie temperatures of the (a) VS$_{2}$, (b) VSe$_{2}$ and (c) VTe$_{2}$ 
monolayers are 292 K, 472 K and 553 K respectively. Our Monte Carlo results regarding the hysteresis 
features of VX$_{2}$ monolayers illustrate that it is possible to observe finite remanence and coercivity 
treatments nearly or well beyond room temperature. These findings also support the previously finding 
Curie temperatures mentioned above. 

\begin{figure}[h!]
\center
\includegraphics[width=7cm]{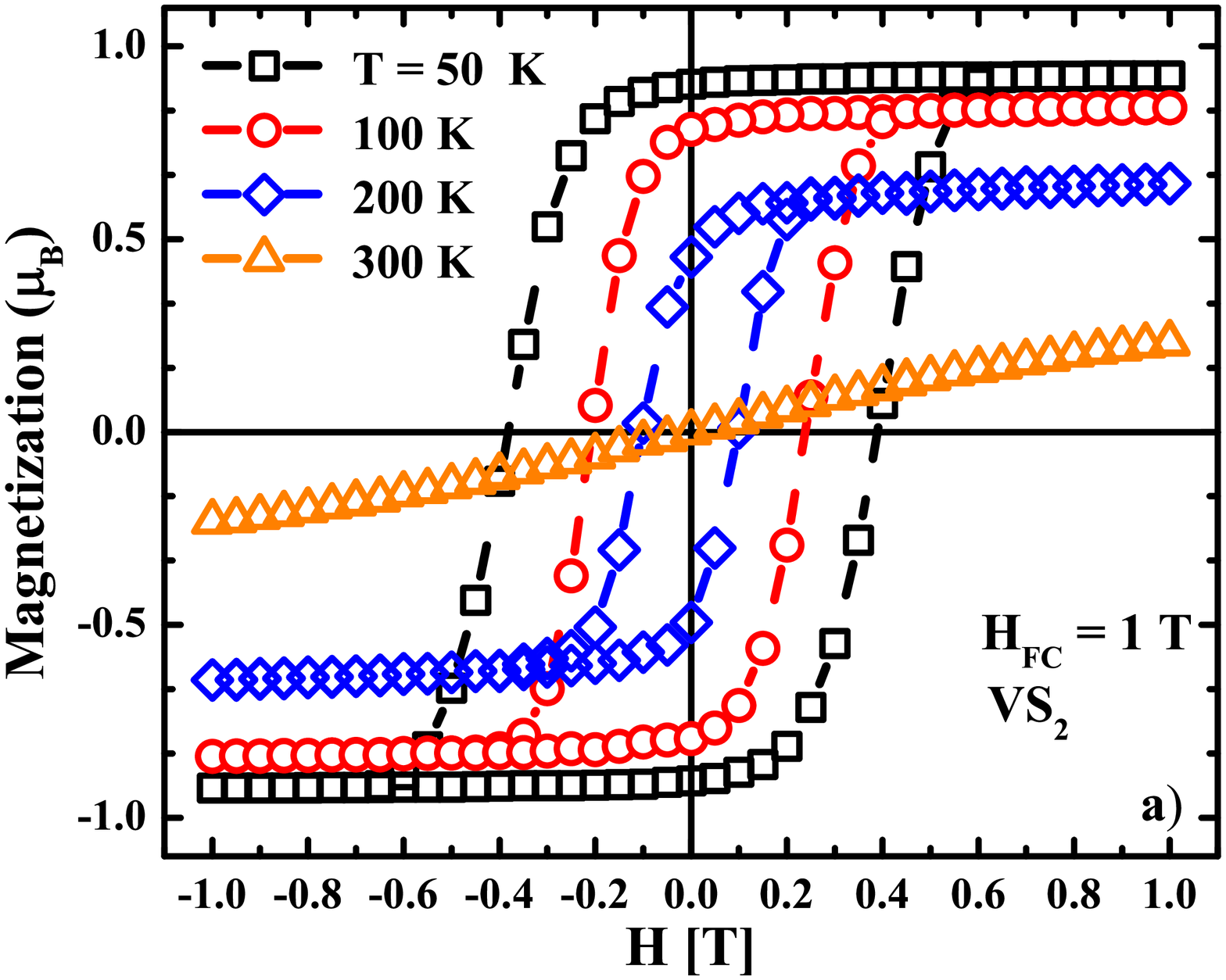}
\includegraphics[width=7cm]{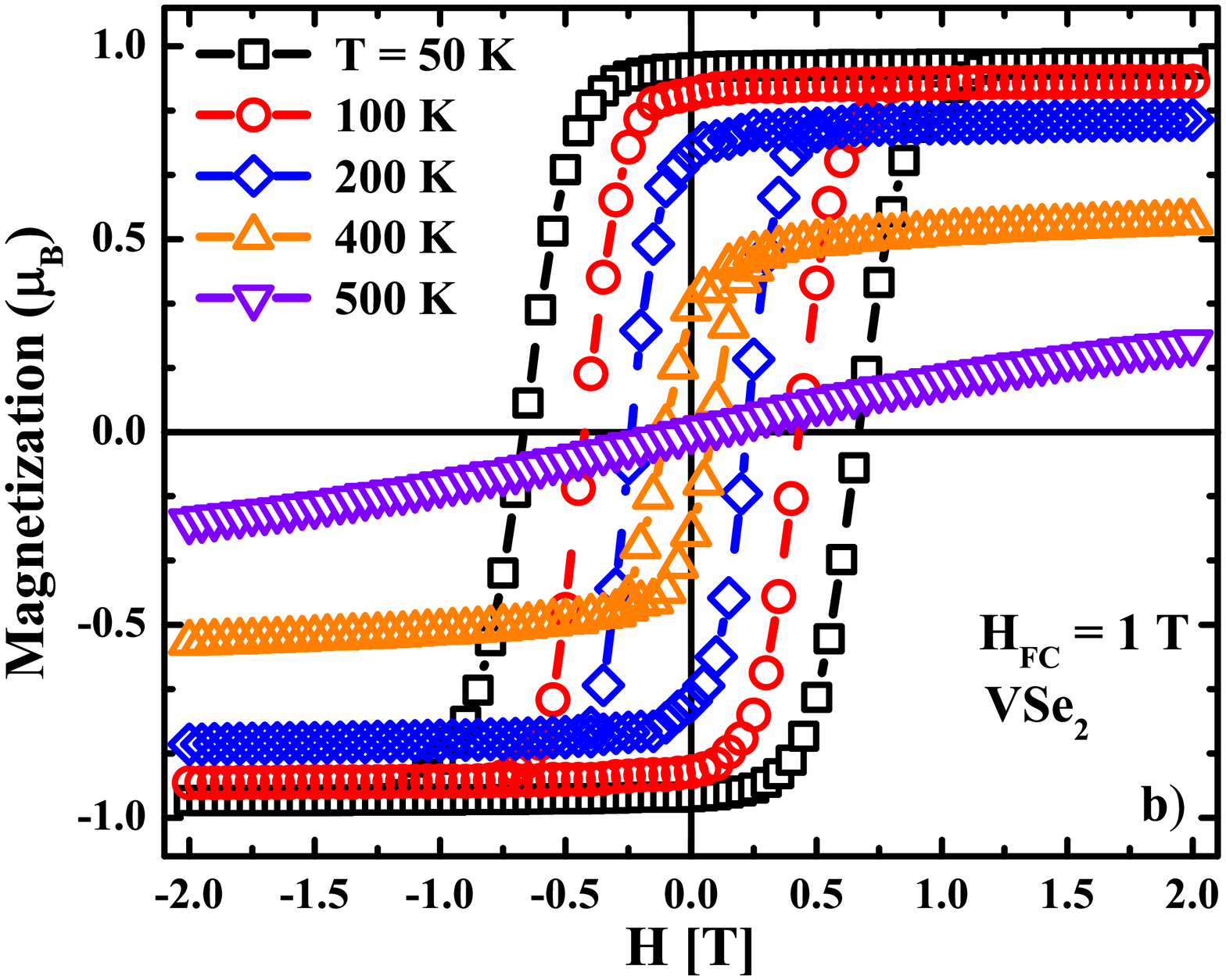}
\includegraphics[width=7cm]{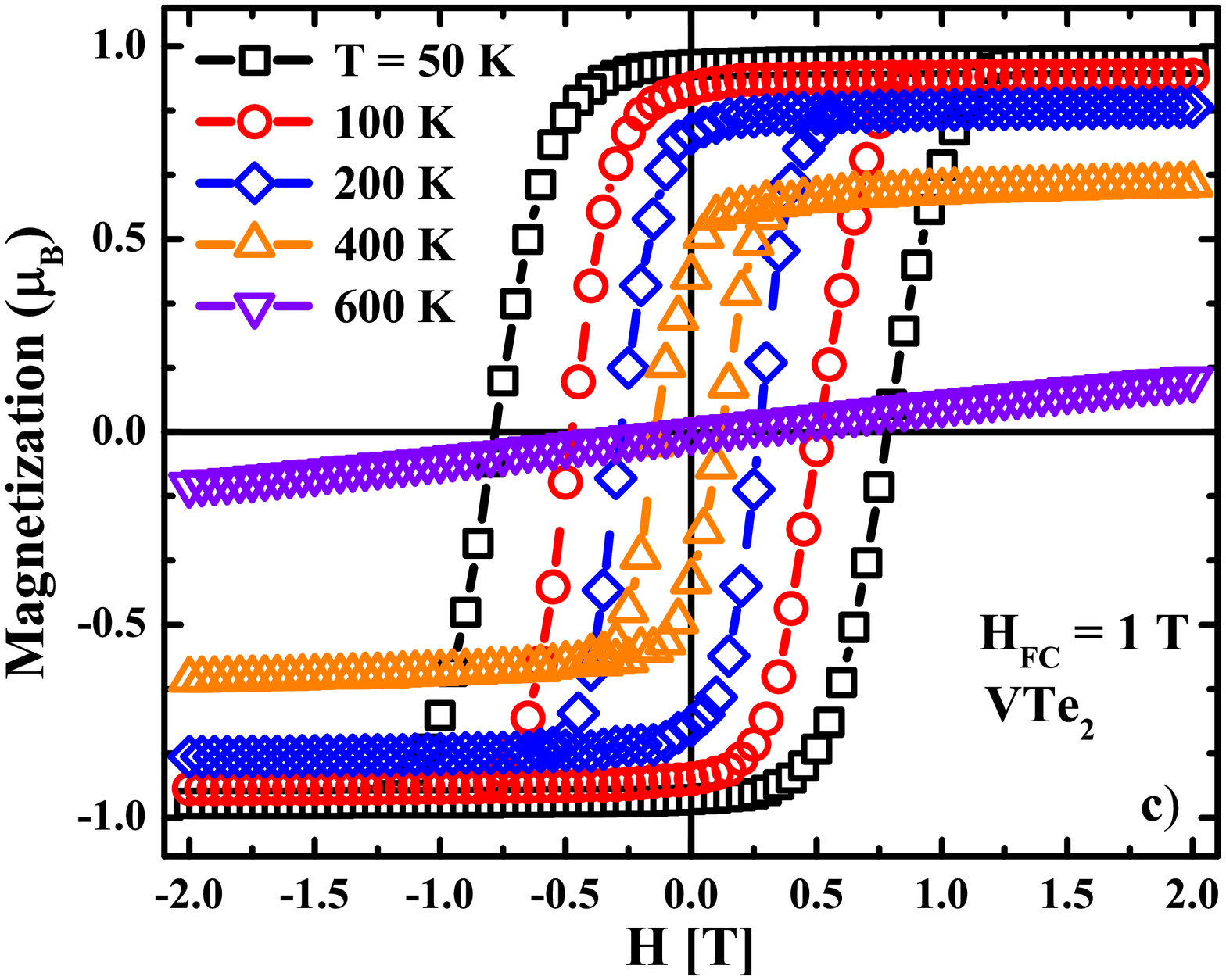}
\caption{(Color online) Effects of the temperature on the hysteresis curves of the (a) VS$_{2}$, (b) VSe$_{2}$ and 
(c) VTe$_{2}$ monolayers. All curves are measured under the field cooling H$_{FC}= 1 T$ applied in 
parallel directions with respect to the VX$_2$ monolayers.}\label{Fig2}
\end{figure}

Final investigation has been devoted to determine the influences of the temperature on the hysteresis 
loops corresponding to remanence and coercivity features depicted in Fig. \ref{Fig1}. In this regard, 
we represent hysteresis curves for (a) VS$_2$, (b) VSe$_2$ and (c) VTe$_2$ monolayers for varying values of the temperature, as  shown in Figs. \ref{Fig2}(a-c). At first glance, shapes of the curves qualitatively resemble to each other. The hysteresis curves measured for all VX$_2$ monolayers clearly reveal the ferromagnetism. In addition to these, shapes of the them explicitly depend on the studied temperature. For example, rectangular-shaped  hysteresis curves obtained  at the relatively lower temperature regions  evolve into S-shape  with increasing temperature. By comparing figures \ref{Fig2}(a), (b) and (c) we observe that VSe$_2$ and VTe$_2$ monolayers evince ferromagnetism above room temperature. 

\section{Concluding Remarks}\label{Conclusion}

In conclusion, we perform a detailed Monte-Carlo simulation based on the Metropolis algorithm to understand the hysteresis properties of the VX$_2$ monolayers. For this aim, we obtain hysteresis curves for a wide range of temperature region up to 600 K. All curves are measured under the field cooling H$_{FC}= 1 T$ applied in  parallel directions with respect to the VX$_2$ monolayers. Results obtain in this study suggest that both coercivity and remanence values sensitively depend on the studied temperature. In other words, they show a tendency to decrease prominently with increasing temperature. Moreover, it is found that hysteresis curves start to evolve from rectangular shaped observed at the lower temperature regions to nearly S-shaped as the temperature is increased further.

To the best of our knowledge, experimental synthesises of VX$_{2}$ monolayers have not been realized up to date.  Hence, we  believe that our Monte Carlo simulation results obtained in this work may be beneficial for both experimental and theoretical point of views.

\section*{Acknowledgements}
The numerical calculations reported in this paper were performed at T\"{U}B\.{I}TAK ULAKBIM (Turkish agency), High Performance and Grid Computing Center (TRUBA Resources).


\end{document}